\documentclass[11pt,a4paper,english,nofootinbib,,superscriptaddress]{revtex4}
\usepackage{lmodern}

\usepackage[T1]{fontenc}
\usepackage[latin9]{inputenc}
\usepackage{babel}
\usepackage{color}

\usepackage{amsmath}
\usepackage{graphicx}
\usepackage{amssymb}
\usepackage{esint}
\usepackage[unicode=true, pdfusetitle,
 bookmarks=true,bookmarksnumbered=false,bookmarksopen=false,
 breaklinks=false,pdfborder={0 0 1},backref=false,colorlinks=false]
 {hyperref}
\usepackage{amsmath}
\usepackage{amssymb}
\def\b{\begin{equation}}
\def\e{\end{equation}}

\def\1{\mbox{I\hspace{-.15em}1}}

\def\R{{\rm I\hspace{-.15em}R}}

\def\b{\begin{equation}}
\def\e{\end{equation}}
\def\bee{\begin{enumerate}}
\def\eee{\end{enumerate}}

\def\R{{\rm I\hspace{-.15em}R}}

\def\bd{\begin{displaystyle}}
\def\ed{\end{displaystyle}}
\def\ba{\begin{array}}
\def\ea{\end{array}}

\def\bee{\begin{enumerate}}
\def\eee{\end{enumerate}}

\def\bes{\begin{eqnarray*}}
\def\ees{\end{eqnarray*}}
\def\be{\begin{eqnarray}}
\def\ee{\end{eqnarray}}

\makeatletter
\@ifundefined{textcolor}{}
{%
 \definecolor{BLACK}{gray}{0}
 \definecolor{WHITE}{gray}{1}
 \definecolor{RED}{rgb}{1,0,0}
 \definecolor{GREEN}{rgb}{0,1,0}
 \definecolor{BLUE}{rgb}{0,0,1}
 \definecolor{CYAN}{cmyk}{1,0,0,0}
 \definecolor{MAGENTA}{cmyk}{0,1,0,0}
 \definecolor{YELLOW}{cmyk}{0,0,1,0}
 }

\usepackage{latexsym}\usepackage{bm}

\makeatother

\begin{document}

\title{ Conformal auxiliary
 "massless" spin-2 field in de Sitter space}

\author{M. Elmizadeh}
\email{elmizadeh@piau.ac.ir } \affiliation{Department of Physics,
Science and Research branch, Islamic Azad University, Tehran,
Iran}

\date{\today}

\begin{abstract}

\noindent \hspace{0.35cm}

Respecting the group theoretical approach,
it is discussed that the linear conformal gravity can be written in
terms of a mixed symmetry tensor field of rank-3 \cite{binegar}. previously, related field equation was obtained in de Sitter space \cite{fatemi}. It was proved
that this kind of tensor field in de Sitter background associates
with two unitary irreducible representations (UIR) of the de
Sitter group. The important fact is that one of the them has a
flat limit, namely, in zero curvature coincides to the UIR
of Poincar\'{e} group which has been studied in Ref \cite{rahbar}.
 However, the second one which is named as auxiliary
field, becomes significant in the study of conformal gravity in de
Sitter background. In this paper we will study the second representation in details by finding its solutions and also the conformally invariant two-point function.

\end{abstract}

\maketitle

\section{Introduction}

Nowadays it is believed that de Sitter (dS) space could potentially describe our universe and the importance dS is that it plays the important role in the inflationary scenario of the early universe \cite{linde}. The isometry group of the dS in four dimensions is SO(1,4), with the same degrees of symmetry as the Minkowskian space. This high level of symmetry of dS may be used as a guideline to study the various kind of field in this curved background which is established by the means of the group theory. This is followed by the fact that in flat space, according to Wigner's interpretation of the elementary systems, one can find a close relationship between the field operators of any kind of spin and the unitary irreducible representation of the Poincar\'e group as the kinematical group of Minkowski space. As a matter of fact the corresponding field equations are established by the Casimir operators of the background isometry group. Extension this fact to dS is straightforward, namely the Casimir operators of the dS group can be used to find the proper field equations where, in the present study using this method, we find the field equations of the massless field of spin-2 in de Sitter space and as it is shown finding the proper solutions and two-point functions are somehow easer that the other methods. Study of the massless spin-2 particle is important because of its
significant role in the quantum cosmology and quantum gravity. In
Minkowski space, the massless field equations are supposed to be
conformal invariant (CI) and for every massless representation of
Poincar\'e group there exists only one corresponding
representation in the conformal group \cite{barut,angelo}. We consider field
equations that the corresponding fields associate with the UIRs of
the de Sitter isometry group SO(1,4). On the other hand being
conformal invariance could be achieved by demanding the field
equations and the solutions should also be transformed according to the
conformal group SO(2,4). In this way the Casimir operators and the
corresponding eigenvalues play the key roles. Previously, Barut
and B\"{o}hm have shown that for the unitary irreducible
representations of conformal group, the eigenvalue of conformal Casimir
operator is $9$ \cite{barut}, this value belongs to a mixed symmetry tensor field of rank-3.  An interesting point arises in the work of Binegar et al \cite{binegar} where they showed that for
any tensor field of rank-2 this value becomes 8. It means that the
tensor field of rank-2 does not correspond to any UIR of conformal
group. This says that, linear conformal quantum gravity must be
formulated in terms of a tensor field of rank-3 and mixed symmetry
with conformal degree zero \cite{binegar}. Mixed symmetry means
that $\Psi_{abc}=-\Psi_{bac},\,\,\,\,\sum_{cycl}\Psi_{abc}=0,$
while a
field of conformal degree zero satisfies $u^d\partial_d\Psi_{abc}=0$.\\
On the other hand according to the group representation theory and its Wigner theorem, a desired CI theory of the spin-2 field should also be invariant under its
space-time symmetry group. So we are interested in extending this group theoretical content to de Sitter space as well as a space-time symmetry group. Quantum field theory in dS  space has developed as a very important subject studied by many authors over the past decade \cite{takook1}. 

It is proved that the theory of massless spin-2 field equations are described by two representation of the de Sitter group in the discrete series which are denoted by 
$\Pi^{\pm}_{2,2}$ and $\Pi^{\pm}_{2,1}$. Only $\Pi^{\pm}_{2,2}$ coincides to the UIR of Poincar\'{e} group, in the flat limit, the other one, $\Pi^{\pm}_{2,1}$, becomes important in the study of conformal gravity. In Ref. \cite{fatemi} in de Sitter background, the conformally invariant field equation for a "massless" spin-2 field of rank-3 was obtained. In this paper we are going to study $\Pi^{\pm}_{2,1}$, which is named as auxiliary fields.

The organization of this paper is as
follows: In section $II$, we introduce the notations of the de Sitter ambient formalism. CI massless wave equations for rank-3 field in dS ambient notation and the corresponding solution is considered in section $III$, and also CI bi-tensor two point function ${\cal W}_{\alpha\beta\gamma \alpha'\beta'\gamma'}(x,x')$  has been calculated in terms
of a conformally coupled scalar two-point function, ${\cal W}_c({\cal Z})$. Finally, in section $V$ the results of the paper are presented for further
investigation. Some useful
mathematical details of calculations are given in the appendix.

\setcounter{equation}{0}

\section{de Sitter space }
De Sitter space-time can be viewed by a 4-dimensional hyperboloid embedded in 5-dimensional flat spacetime:
\begin{equation}
X_H=\{x \in \R^5 ;x^2=\eta_{\alpha\beta} x^\alpha x^\beta
=-H^{-2}=-\frac{3}{\Lambda}\},\;\;\alpha,\beta=0,1,2,3,4,
\end{equation}where $\eta_{\alpha\beta}=$ diag$(1,-1,-1,-1,-1)$ and $H$,
$\Lambda$ are the Hubble parameter and cosmological constant
respectively. The dS metric is
$$ds^2=\eta_{\alpha\beta}dx^{\alpha}dx^{\beta}\mid_{x^2=-H^{-2}}=g_{\mu\nu}^{dS}dX^{\mu}dX^{\nu},\;\;\mu,\nu=0,1,2,3$$
where $X^\mu$'s are  $4$ space-time intrinsic coordinates of the dS
hyperboloid. Any geometrical object in this space can be written
in terms of both four local coordinates $X^\mu$ (intrinsic space
notation) and five global coordinates $x^\alpha$ (ambient space
notation). Kinematical group of the de Sitter space is the 10-parameter group $SO_0(1,4)$ (connected component of the identity in $O(1,4)$ ), which is one of the two possible deformations of the Poincar\'{e} group. For simplicity, one can put $H=1$.\\
In the following the ambient space notation is used; in these notation, the relationship with unitary irreducible representations of de Sitter group becomes straightforward because the Casimir operators are easily identifiable \cite{gazeau}. There are two Casimir operators (Appendix)
\begin{equation}\label{q3} Q^{(1)}_3=-\frac{1}{2}L^{\alpha\beta}L_{\alpha\beta},\;\;\;\;\;
Q^{(2)}_3=-W_{\alpha}W^\alpha,\end{equation} where $W_\alpha=-\frac{1}{8}\epsilon_{\alpha\beta\gamma\sigma\eta}L^{\beta\gamma}L^{\sigma\eta},$ with $10$ infinitesimal generator $L_{\alpha\beta}=M_{\alpha\beta}+S_{\alpha\beta}$ and the symbol $\epsilon_{\alpha\beta\gamma\sigma\eta}$ holds for the usual antisymmetric tensor. The subscript $3$ in $Q^{(1)}_3$ and $Q^{(2)}_3$ reminds us that the carrier space is constituted by third rank tensors. The orbital part $M_{\alpha\beta}$, and the action of the spinorial part $S_{\alpha\beta}$ on a rank-3 tensor field $F_{\alpha\beta\gamma}$ are defined on the ambient space, read respectively \cite{gazeau1}
$$M_{\alpha\beta}=-i(x_\alpha\partial_\beta-x_\beta\partial_\alpha)=-i(x_\alpha\Bar\partial_\beta-x_\beta\Bar\partial_\alpha),$$
\begin{equation}S_{\alpha\beta}F_{\gamma\delta\sigma}=-i(\eta_{\alpha\gamma}F_{\beta\delta\sigma}-\eta_
{\beta\gamma}F_{\alpha\delta\sigma}+\eta_{\alpha\delta}F_{\gamma\beta\sigma}-\eta_{\beta\delta}F_{\gamma\alpha\sigma}+\eta_{\alpha\sigma}F_{\gamma\delta\beta}-\eta_{\beta\sigma}F_{\gamma\delta\alpha}).\end{equation}
$\Bar\partial_\alpha$ is the tangential (or transverse) derivative on dS space defined by \begin{equation}\label{ar}\Bar\partial_\alpha=\theta_{\alpha\beta}\partial^\beta=\partial_\alpha+x_\alpha x\cdot{\partial},\;\;\;\;x\cdot{\Bar\partial}=0,\end{equation}
and $\theta_{\alpha\beta}=\eta_{\alpha\beta}+x_\alpha x_\beta$ is the transverse projector.

As we know the operator $Q^{(1)}_3$ commutes with the action of the group generators and, so it is constant in each UIR. Thus the eigenvalues of $Q^{(1)}_3$ can be used to classify UIR \textit{i.e.,}\begin{equation}(Q^{(1)}_3-\langle{Q^{(1)}_3}\rangle)F(x)=0.\end{equation}

 We have a classification scheme using a pair $(p,q)$ of parameters involved in the following possible spectral values of the Casimir operator \cite{dixmier}:\begin{equation}Q^{(1)}_p=(-p(p+1)-(q+1)(q-2))I_d,\;\;\;\;\;\;\;\;Q^{(2)}_p=(-p(p+1)q(q-1))I_d.
\end{equation}

Three types of scaler, tensorial or spinorial UIR are distinguished for $SO(1,4)$ according to the range of values of the parameters $q$ and $p$ \cite{dixmier,takahashi}, namely: the principal, the complementary and the discrete series. The spin-2 tensor representations are in the following relations:
\begin{itemize}
\item[1)]The UIR $U^{2,\nu}$
in the principal series where $p=s=2$ and $q=\frac{1}{2}+i\nu$ correspond to the Casimir spectral values:
\begin{equation}\langle{Q^{(\nu)}_3}\rangle=\nu^2-\frac{15}{4},\;\;\;\;\nu \in R,\end{equation} note that $U^{2,\nu}$ and $U^{2,\-nu}$ are equivalent.

\item[2)]The UIR $V^{2,q}$
in the complementary series where $p=s=2$ and $q-q^2=\mu$, correspond to \begin{equation}\langle{Q^{(\mu)}_3}\rangle=q-q^2-4\equiv\mu-4,\;\;\;\;\ 0<\mu<\frac{1}{4}.\end{equation}
\item[3)]The UIR $\Pi^{\pm}_{2,q}$
in the discrete series where $p=s=2$ correspond to \begin{equation}\langle{Q^{(1)}_3}\rangle=-4,\;\;\;\;\;q=1 (\Pi^{\pm}_{2,1}) \;\;\;\;\;\langle{Q^{(2)}_3}\rangle=-6,\;\;\;\;\;q=2 (\Pi^{\pm}_{2,2}), \end{equation}
\end{itemize}
The "massless" spin-2 field in de Sitter spactime corresponds to the $\Pi^{\pm}_{2,2}$ and $\Pi^{\pm}_{2,1}$ cases, in which the sign $\pm$, stands for the helicity of the representation.
The $\Pi^{\pm}_{2,2}$, in the discrete series with $p=q=2$, have a Minkowskian interpretation. It is important to note that $p$ and $q$ do not bear the meaning of mass and spin, and for discrete series in flat limit, $p=q=s$ are indeed none other than spin. It is necessary to explain about concept of mass in dS space, the term "massless" is referred to fields that propagate on light-cone and the term "massive" is referred to field that in their zero curvature limit would be  reduced to massive Minkowskian fields \cite{barut}. Concept of light-cone propagation, however, does exist and leads to the conformal invariance. The representation $\Pi^+_{2,2}$
has a unique extension to a direct sum of two representations $C(3;2,0)$ and
$C(-3;2,0)$ of the conformal group, with positive and negative
energies respectively \cite{barut,levy}. The latter restricts to
the massless Poincar\'e UIR $P^>(0, 2)$ and $P^<(0,2)$ with
positive and negative energies respectively. $ {\cal P}^{
\stackrel{>} {<}}(0,2)$ and $(resp.{\cal
P}^{\stackrel{>}{<}}(0,-2))$  are the massless Poincar\'e representations
with positive and negative energies and positive (resp. negative)
helicity. The following diagrams illustrate these connections
\begin{equation}
\left.
\begin{array}{ccccccc}
&& {\cal C}(3,2,0)& &{\cal C}(3,2,0)&\hookleftarrow &{\cal P}^{>}(0,2)\\
\Pi^+_{2,2} &\hookrightarrow &\oplus&\stackrel{H=0}{\longrightarrow} & \oplus  & &\oplus\\
&& {\cal C}(-3,2,0)& & {\cal C}(-3,2,0) &\hookleftarrow &{\cal
P}^{<}(0,2),\\
\end{array}
\right.
\end{equation}

\begin{equation}
\left.
\begin{array}{ccccccc}
&& {\cal C}(3,0,2)& &{\cal C}(3,0,2)&\hookleftarrow &{\cal P}^{>}(0,-2)\\
\Pi^-_{2,2} &\hookrightarrow &\oplus&\stackrel{H=0}{\longrightarrow}&\oplus &&\oplus\\
&& {\cal C}(-3,0,2)&& {\cal C}(-3,0,2)&\hookleftarrow &{\cal P}^{<}(0,-2),\\
\end{array}
\right.
\end{equation} where the arrows $\hookrightarrow $ designate unique
extension. Note that the representations
$\Pi^{\pm}_{2,1}$ do not have corresponding flat limit.
Mathematical details of the group contraction and the physical
principles underlying the relationship between dS and Poincar\'e
groups can be found in Ref.s \cite{levy} and \cite{bacry},
respectively.

\setcounter{equation}{0}

\section{Conformally invariant field equations}

The equations of massless fields are CI, a trivial example is a photon field which satisfies conformally invariant Maxwell equation \cite{maxwell}. In this way the field equations for a massless spin-2 tensor field should be CI. To view this fact from the group theory one knows that the conformal group acts non-linearly on Minkowski coordinates, so that preserving conformal invariance property is difficult. A rather simple method proposed by Dirac who presented a manifestly conformally covariant formulation in
which the Minkowski coordinates are replaced by coordinates on
which the conformal group acts linearly. In this method the theory is formulated on a 5-dimensional hyper-cone in a 6-dimensional space that named Dirac's six cone \cite{dirac}. Within
this formalism, the scalar, spinor and vector
conformally invariant fields in $d = 4$ were obtained by him. This theory
developed in some papers  \cite{Mack, mack, kastrup}, the
generalization to dS space was done in \cite{gareta1, ta4} to obtain CI
field equations for scalar, vector and symmetric rank-2 tensor fields. Relegating the mathematical details to the references, let us write the CI wave equation for the tensor field of rank-3 from Ref. \cite{fatemi} 
$$(Q_3+6)(Q_3+4)F_{\alpha\beta\gamma}=0$$ where $Q_3$ is the Casimir operator of dS group (it is introduced by (\ref{q3})). This equation can be considered as two equation of order two \\
\mbox{$(Q_3+6)F_{\alpha\beta\gamma}=0$,\;\;\;transforms according to $\Pi^{\pm}_{2,2}$,}\\
\mbox{$(Q_3+4)F_{\alpha\beta\gamma}=0$,\;\;\;transforms according to $\Pi^{\pm}_{2,1}$.}\\
In the flat limit $(H\rightarrow0)$, $\Pi^{\pm}_{2,2}$ coincides to the UIR of the Poincar\'{e} group, but $\Pi^{\pm}_{2,1}$ does not have any corresponding Poincar\'{e} representation, however both representations become important in the study of the CI field equation in dS space. Studying the solutions corresponding to $\Pi^{\pm}_{2,2}$ has been done before \cite{rahbar}, but in this paper we consider the fields which transform according to $\Pi^{\pm}_{2,1}$, here, the fields are named as auxiliary fields, since there is no flat limit to them.\\

\subsection*{de Sitter field solution}

In this section we find the solution of the following field equation
\begin{equation}\label{br6}(Q_0-2)F_{\alpha\beta\gamma}=0\;\;\;\;or\;\ (Q_3+4)F_{\alpha\beta\gamma}=0,\end{equation}
The general solution of the above equation can be written as \cite{rahbar}
\begin{equation}\label{br7}F_{\alpha\beta\gamma}=(\Bar\partial_\alpha+x_\alpha){\cal K}_{\beta\gamma}-(\Bar\partial_\beta+x_\beta){\cal K}_{\alpha\gamma}+\bar{Z}_\alpha{\cal H}_{\beta\gamma}-\bar{Z}_\beta{\cal H}_{\alpha\gamma},\end{equation} where ${\cal K}_{\alpha\beta}$ and ${\cal H}_{\alpha\beta}$ are two tensor fields of rank-2, $Z$ is a constant 5-dimensional vector and bar over this vector guarantees the transversality (see$(\ref{ar}$).  Imposing the transversality, divergencelessness and tracelessness conditions on $F_{\alpha\beta\gamma},$ leads one to write 

$${\cal K}_{\beta\gamma}= \Big(-\frac{1}{2}(x\cdot{Z})+\frac{1}{8}(Z\cdot{\Bar\partial})\Big){\cal H}_{\beta\gamma}-\frac{1}{8}\Big(x_\beta Z\cdot{{\cal H}_{.\gamma}}+x_\gamma Z\cdot{{\cal H}_{\beta.}}\Big)
$$
\begin{equation}{\cal H}_{\alpha\beta}(x)={\cal D}_{\alpha\beta}(x,\partial,Z_1,Z_2)\phi,\end{equation} where
\begin{equation}{\cal D}(x,\partial,Z_1,Z_2)= \Big(-\frac{2}{3}\theta Z_{1.}+{\cal S}\bar{Z}_1+\frac{1}{3}D_2\Big[\frac{1}{9}D_1Z_{1.}+x\cdot{Z_1}\Big]\Big) \Big(\bar{Z}_2+D_1(x\cdot{Z_2})\Big),\end{equation}
where ${\cal S}$ is symmetrizer operator, $D_1=\Bar\partial$, $D_{2}={\cal S}(\Bar\partial-x)$ and $Z_1,$ $Z_2$ are another two 5-dimensional constant vectors and $\phi$ is a massless conformally coupled scalar field. The important fact is that we could write the solution in terms of a massless scalar filed in de Sitter space. In what follows we will find the two-point function

\subsection*{two-point function}

We find the related the two-point functions in dS space in terms
of the bi-tensors \cite{allen2}. Actually the bi-tensors are functions of two points
$(x, x')$ which behave like tensors under each coordinate. Bi-tensors are called maximally
symmetric if they respect to the de Sitter invariance. Furthermore, as
explained in \cite{bros} and \cite{garidigarou}, the axiomatic
field theory in de Sitter is based on bi-tensor Wightman two-point
function. This two-point function is defined by \begin{equation} {\cal{W}}_{\alpha\beta \gamma\alpha'\beta'\gamma'}(x,x')=\langle
\Omega|F_{\alpha\beta\gamma}(x)F_{\alpha'\beta'\gamma'}(x')|\Omega  \rangle,\end{equation} where $x,x'\in X_H$ and $|\Omega\rangle $ is the Fock-vacuum
state. The two-point function which is a solution of the Eq. (\ref{br6}) with respect to $x$ and $x'$, can be written in terms of transverse two-point functions, ${\cal{W}}^{\cal{K}}_{\beta\gamma\beta'\gamma'}$ and ${\cal{W}}^{\cal{H}}_{\beta\gamma\beta'\gamma'}$, which will be identified later. We have
\begin{equation}\label{br8}{\cal W}_{\alpha\beta \gamma\alpha'\beta'\gamma'}(x,x')= \bar{{\cal S}}_{\alpha\beta}(\Bar\partial_\alpha+x_\alpha)\Big(\bar
{{\cal S }}'_{\alpha'\beta'}(\Bar\partial'_{\alpha'}+{x'}_{\alpha'}){\cal W}^{\cal K}_{\beta\gamma\beta'\gamma'}(x,x')\Big)+\bar{{\cal S}}_{\alpha\beta}
\bar{{\cal S}}'_{\alpha'\beta'}\Big((\theta_\alpha\cdot{{\theta'}_{\alpha'}}){\cal W}^{\cal H}_{\beta\gamma\beta'\gamma'}(x,x')\Big),\end{equation}
where anti-symmetrizer operator is defined by $\bar{{\cal S}}_{\alpha\beta}{\cal K}_{\alpha\beta}\equiv{\cal K}_{\alpha\beta}-{\cal K}_{\beta\alpha}$ and $\bar{{\cal S}}'_{\alpha'\beta'}{\cal K}_{\alpha'\beta'}\equiv{\cal K}_{\alpha'\beta'}-{\cal K}_{\beta'\alpha'}$ which we use them to get shorthand equations.
After some calculation we get
\begin{equation}\begin{aligned}
\label{br9}\bar{{\cal S}}'_{\alpha'\beta'}(\Bar\partial'_{\alpha'}+{x'}_{\alpha'}){\cal W}^{\cal K}_{\beta\gamma\beta'\gamma'}(x,x')=\bar{{\cal S}}'_{\alpha'\beta'}\Big(-\frac{1}{2}(x\cdot{{\theta'}_{\alpha'}})+\frac{1}{8}({\theta'}_{\alpha'}\cdot{\Bar\partial})\Big){\cal W}^{\cal H}_{\beta\gamma\beta'\gamma'}(x,x')\\-\frac{1}{8}\bar{{\cal S}}'_{\alpha'\beta'}\Big(x_\beta{\theta'}_{\alpha'}\cdot{{\cal{W}}^{\cal{H}}_{.\gamma\beta'\gamma'}}(x,x')+x_\gamma{\theta'}_{\alpha'}\cdot{{\cal{W}}^{\cal{H}}_{\beta .\beta'\gamma'}}(x,x')\Big).
\end{aligned}\end{equation}
and also ${\cal W}^{\cal H}_{\beta\gamma\beta'\gamma'}$ was derived in \cite{pejhan}:
\begin{equation}
\begin{array}{l}{\cal W}^{\cal H}_{\beta\gamma\beta'\gamma'}(x,x')=-\frac{2}{3}{\cal S}'\theta\theta'\cdot{(\theta\cdot{\theta'}+D_1(\theta'\cdot{x}))}+{\cal S}{\cal S}'\theta\cdot{\theta'}(\theta\cdot{\theta'}+D_1(\theta'\cdot{x}))\\+\frac{1}{3}D_2{\cal S}'\Big(\frac{1}{9}D_1\theta'\cdot+x\cdot{\theta'}\Big)(\theta\cdot{\theta'}+D_1(\theta'\cdot{x})){\cal W}_c(x,x'),\end{array}
\end{equation}

${\cal W}_c(x,x')$ is the two-point function for a conformally coupled scalar field \cite{chernikov}:
\begin{equation}{\cal{W}}_c(x,x')=-\frac{1}{8\pi^2}\Big[{\cal{P}}\frac{1}{1-{\cal Z}(x,x')}-i\pi\epsilon(x^0-x'^0)\delta(1-{\cal Z}(x,x'))\Big],\end{equation}
with
\begin{equation} \epsilon (x^0-x'^0)=\left\{ \ba{rcl} 1&x^0>x'^0 ,\\
0&x^0=x'^0 ,\\ -1&x^0<x'^0.\\ \ea\right.\end{equation}
In some cases of interest it is useful to express the relations in
terms of $ {\cal{Z}}$ which is an invariant object under the
isometry group $O(1,4)$. It is defined for two given points on the
dS hyperboloid $x$ and $x'$, by: $$ {\cal{Z}}\equiv
-x.x'=1+{\frac{1}{2}}(x-x')^2,$$ note that any function of ${\cal
Z}$ is dS-invariant, as well. It is the work of a lot of lines to
show that (\ref{br8}) in terms of $ {\cal Z}$ becomes:

\begin{equation}
\begin{array}{l}{\cal W}_{\alpha\beta \gamma\alpha'\beta'\gamma'}(x,x')=\frac{1}{216}\bar{{\cal S}}'_{\alpha'\beta'}\bar{{\cal S}}_{\alpha\beta}{\cal S}'_{\beta'\gamma'}{\cal S}_{\beta\gamma}\Big[\theta_{\beta\gamma}{\theta'}_{\beta'\gamma'}(\theta_\alpha\cdot{{\theta'}_{\alpha'}})\Big(-80+57{\cal Z}\frac{d}{d{\cal Z}}\Big)+\\
x_\beta\theta_{\alpha\gamma}{\theta'}_{\beta'\gamma'}(x\cdot{{\theta'}_{\alpha'}})\Big(64-34{\cal Z}\frac{d}{d{\cal Z}}\Big)+
(x'\cdot\theta_\alpha)(x\cdot{{\theta'}_{\alpha'}})\theta_{\beta\gamma}{\theta'}_{\beta'\gamma'}\Big(\frac{68{\cal Z}+(-30-114{\cal Z}^2)\frac{d}{d{\cal Z}}}{1-{\cal Z}^2}\Big)+\\
{\theta'}_{\beta'\gamma'}(\theta_\alpha\cdot{{\theta'}_{\alpha'}})(x'\cdot{\theta_\beta})(x'\cdot{\theta_\gamma})\Big(\frac{-186-276{\cal Z}\frac{d}{d{\cal Z}}}{1-{\cal Z}^2}\Big)+
x_\beta{\theta'}_{\beta'\gamma'}(x\cdot{{\theta'}_{\alpha'}})(x'\cdot{\theta_\alpha})(x'\cdot{\theta_\gamma})\Big(\frac{-68+140{\cal Z}\frac{d}{d{\cal Z}}}{1-{\cal Z}^2}\Big)+\\
(\theta_\alpha\cdot{{\theta'}_{\alpha'}})(\theta_\beta\cdot{{\theta'}_{\beta'}})(\theta_\gamma\cdot{{\theta'}_{\gamma'}})\Big(\frac{(152-156{\cal Z}^2)+(7{\cal Z}-31{\cal Z}^3)\frac{d}{d{\cal Z}}}{1-{\cal Z}^2}\Big)+\\
x_\beta(x\cdot{{\theta'}_{\alpha'}})(\theta_\alpha\cdot{{\theta'}_{\beta'}})(\theta_\gamma\cdot{{\theta'}_{\gamma'}})\Big(\frac{(-456+452{\cal Z}^2)+(-9{\cal Z}-15{\cal Z}^3)\frac{d}{d{\cal Z}}}{1-{\cal Z}^2}\Big)+\\
\theta_{\alpha\beta}(x\cdot{{\theta'}_{\alpha'}})(x\cdot{{\theta'}_{\beta'}})(\theta_\gamma\cdot{{\theta'}_{\gamma'}})\Big(\frac{(-880+1960{\cal Z}^2-938{\cal Z}^4)+(-278{\cal Z}-700{\cal Z}^3-374{\cal Z}^5)\frac{d}{d{\cal Z}}}{(1-{\cal Z}^2)^2}\Big)+\\
(x\cdot{{\theta'}_{\alpha'}})(x'\cdot{\theta_\alpha})(\theta_\gamma\cdot{{\theta'}_{\gamma'}})(\theta_\beta\cdot{{\theta'}_{\beta'}})\Big(\frac{(-38{\cal Z}+94{\cal Z}^3)+(-135+710{\cal Z}^2-427{\cal Z}^4)\frac{d}{d{\cal Z}}}{(1-{\cal Z}^2)^2}\Big)+\\
x_\alpha(x\cdot{{\theta'}_{\alpha'}})(x\cdot{{\theta'}_{\gamma'}})(x'\cdot{\theta_\gamma})(\theta_\beta\cdot{{\theta'}_{\beta'}})\Big(\frac{(-74{\cal Z}+50{\cal Z}^3)+(300-760{\cal Z}^2+412{\cal Z}^4)\frac{d}{d{\cal Z}}}{(1-{\cal Z}^2)^2}\Big)+\\
x_\alpha\theta'_{\beta'\gamma'}(\theta'_{\alpha'}\cdot{\theta_\beta})(x\cdot{\theta_\gamma})\Big(75\frac{d}{d{\cal Z}}\Big)+\\
(x\cdot{{\theta'}_{\beta'}})(x'\cdot{\theta_\beta})(x\cdot{{\theta'}_{\gamma'}})(x'\cdot{\theta_\gamma})(\theta_\alpha\cdot{{\theta'}_{\alpha'}})\Big(\frac{(-526+1592{\cal Z}^2-874{\cal Z}^4)+(-806{\cal Z}+2862{\cal Z}^3-1648{\cal Z}^5)\frac{d}{d{\cal Z}}}{(1-{\cal Z}^2)^3}\Big)+\\
(\frac{(208-112{\cal Z}^2)+(464{\cal Z}-272{\cal Z}^3)\frac{d}{d{\cal Z}}}{1-{\cal Z}^2}\Big)\Big]{\cal W}_c({\cal Z}).
\end{array}
\end{equation}
\setcounter{equation}{0}

\section{Conclusion}

It was shown that linear conformal gravity may be understood by the means of the group theory in de Sitter and flat backgrounds, where the isometry group Poinca\'e of de Sitter groups play the role of the space time symmetry group. As one knows, being the conformal invariance may also be understood as the invariance of the conformal group. This imposes a constraint on the field equations and the solutions. Namely it is proved that only a mixed symmetric tensor field of rank three can potentially describe the gravity \cite{binegar}. Previously, in Ref. \cite{fatemi} CI field equation for this tensor field of rank-3 in dS space by the means of the Dirac six-cone formalism has been obtained as following
$$(Q_3+4)(Q_3+6)F_{\alpha\beta\gamma}=0.$$
This field equation leads to two equations: $(Q_3+6)F_{\alpha\beta\gamma}=0$ which has been studied in \cite{rahbar}, and  $(Q_3+4)F_{\alpha\beta\gamma}=0$, which studied in this paper. We found the solutions also two-point function, these are found in terms of the massless conformally coupled scalar two point function. It was shown that the two-point function is dS and conformal invariant, and it may be used in obtain quantum gravity two point function.


\setcounter{equation}{0}

\begin{appendix}

\section{Some useful relations}
In this appendix, some useful relations are given. The action of Casimir operators of de Sitter group $Q_1$ (vector Casimir operator), $Q_2$ and $Q_3)$ (Casimir operator for the rank-2 and rank-3 tensor field respectively ) can be written in the more explicit form \cite{tanhayi}
\begin{equation}Q_1K_\alpha=(Q_0-2)K_\alpha+2x_\alpha\partial\cdot{K}-2\partial_\alpha x\cdot{K},\end{equation}
\begin{equation}Q_2{\cal K}_{\alpha\beta}=(Q_0-6){\cal K}_{\alpha\beta}+2{\cal S}x_\alpha\partial\cdot{{\cal K}_\beta}-2{\cal S}\partial_\alpha x\cdot{{\cal K}_\beta}+2\eta_{\alpha\beta}{\cal K}',\end{equation}where ${\cal K}'={\cal K}_\alpha^\alpha$.
\begin{equation}\begin{aligned}
Q_3F_{\alpha\beta\gamma}=(Q_0-6) F_{\alpha\beta\gamma}+2 \Big(\eta_{\alpha\beta} F_{..\gamma}+ \eta_{\alpha\gamma}F_{.\beta.}+ \eta_{\beta\gamma} F_{\alpha..}\Big)+2\Big(x_{\alpha}\Bar \partial\cdot {F_{.\beta\gamma}}+\\
x_{\beta}\Bar\partial\cdot{F_{\alpha.\gamma}}+x_{\gamma}\Bar \partial\cdot {F_{\alpha\beta.}}\Big)-2\Big(\partial_{\alpha}x\cdot{F_{.\beta\gamma}}+\partial_{\beta}x\cdot{F_{\alpha.\gamma}}+\partial_{\gamma}x\cdot{F_{\alpha\beta.}}\Big)-2\Big(F_{\alpha\gamma\beta}+ F_{\beta\alpha\gamma}+F_{\gamma\beta\alpha} \Big),
 \end{aligned}\end{equation}
 where $F_{..\gamma}=F_{\alpha\alpha\gamma}.$ After imposing mixed symmetry condition on $F_{\alpha\beta\gamma},$ above relation becomes
 \begin{equation}\begin{aligned}
Q_3F_{\alpha\beta\gamma}=(Q_0-4) F_{\alpha\beta\gamma}-2F_{\beta\gamma\alpha}-2F_{\gamma\alpha\beta}+2 \Big(\eta_{\alpha\beta} F_{..\gamma}+ \eta_{\alpha\gamma}F_{.\beta.}+ \eta_{\beta\gamma} F_{\alpha..}\Big)+2\Big(x_{\alpha}\Bar \partial\cdot {F_{.\beta\gamma}}+\\
x_{\beta}\Bar\partial\cdot{F_{\alpha.\gamma}}+x_{\gamma}\Bar \partial\cdot {F_{\alpha\beta.}}\Big).
\end{aligned}\end{equation}

To obtain the two-point function, the following identities are used

\begin{equation}\Bar\partial_\alpha f({\cal Z})=-(x'\cdot{\theta_\alpha})\frac{df({\cal Z})}{d({\cal Z})},\end{equation}
\begin{equation}\theta^{\alpha\beta}\theta'_{\alpha\beta}=\theta\cdot\cdot{\theta'}=3+{\cal Z}^2,\end{equation}
\begin{equation}(x\cdot{\theta'_{\alpha'}})(x\cdot{\theta'^{\alpha'}})={\cal Z}^2-1,\end{equation}
\begin{equation}(x\cdot{\theta'_\alpha})(x'\cdot{\theta^\alpha})={\cal Z}(1-{\cal Z}^2),\end{equation}
\begin{equation}\Bar\partial_\alpha(x\cdot{\theta'^{\beta'}})=\theta_\alpha\cdot{\theta'^{\beta'}},\end{equation}
\begin{equation}\Bar\partial_\alpha(x'\cdot{\theta_\beta})=x_\beta(x'\cdot{\theta_\alpha})-{\cal Z}\theta_{\alpha\beta},\end{equation}
\begin{equation}\Bar\partial_\alpha(\theta_\beta\cdot{\theta'_{\beta'}})=x_\beta(\theta_\alpha\cdot{\theta'_{\beta'}})+ \theta_{\alpha\beta}(x\cdot{\theta'_{\beta'}}),\end{equation}
\begin{equation}\theta'^{\beta}_{\alpha'}(x'\cdot{\theta_\beta})=-{\cal Z}(x\cdot{\theta'^{\alpha'}}),\end{equation}
\begin{equation}\theta'^{\gamma}_{\alpha'}(\theta_\gamma\cdot{\theta'_{\beta'}})=\theta'_{\alpha'\beta'}+(x\cdot{\theta'^{\alpha'}})(x\cdot{\theta'^{\beta'}}),\end{equation}
\begin{equation}Q_0f({\cal Z})=(1-{\cal Z}^2)\frac{d^2f({\cal Z})}{d{\cal Z}^2}-4{\cal Z}\frac{df({\cal Z})}{d({\cal Z})}.\end{equation}
\end{appendix}

\end{document}